# A machine learning approach to predict the structural and magnetic properties of Heusler alloy families

Srimanta Mitra* [a,b], Aquil Ahmad [c], Sajib Biswas [a] and Amal Kumar Das*[a]

[a] Indian Institute of Technology Kharagpur, Khargapur-721302, India

[b] Sensor Development Area, Space Applications Centre, ISRO, Ahmedabad-380015, India

[c] National Changhua University of Education, Changhua city, Taiwan-500

*email: [a, b] srimanta_44@sac.isro.gov.in; [a] amal@phy.iitkgp.ac.in;

**Abstract**

Random forest (RF) regression model is used to predict the lattice constant, magnetic moment and formation energies of full Heusler alloys, half Heusler alloys, inverse Heusler alloys and quaternary Heusler alloys based on existing as well as indigenously prepared databases. Prior analysis was carried out to check the distribution of the data points of the response variables and found that in most of the cases, the data is not normally distributed. The outcome of the RF model performance is sufficiently accurate to predict the response variables on the test data and also shows its robustness against overfitting, outliers, multicollinearity and distribution of data points. The parity plots between the machine learning predicted values against the computed values using density functional theory (DFT) shows linear behavior with adjusted $R^2$ values lying in the range of 0.80 to 0.94 for all the predicted properties for different types of Heusler alloys. Feature importance analysis shows that the valence electron numbers plays an important feature role in the prediction for most of the predicted outcomes. Case studies with one full Heusler alloy and one quaternary Heusler alloy were also mentioned comparing the machine learning predicted results with our earlier theoretical calculated values and experimentally measured results, suggesting high accuracy of the model predicted results.

1. Introduction

In recent years, there is a constant increasing interest in the field of machine learning (ML) methods in multiple areas of research, e.g. image recognition [1-3], speech recognition [4-5], event forecasting [6-8], pattern recognition [9], self-driving cars [10-11], medical diagnosis [12-13], automatic language translation [14-15] etc. Materials informatics (MI) is one of the emerging areas of study where informatics is used to understand and predict materials' properties [16-18]. Applications of ML in material science domain is much attractive and is of keen interest in current time [19-25]. In MI, ML models are used to develop and predict new materials' properties using different databases [26-30, 31]. There are many studies that have been carried out to predict the electronic band gap [32-33], magnetism [34-36], magnetic entropy change [37-38], superconducting critical temperature [39-43], melting temperatures [44-45], dielectric properties [46-47] , molecular atomization energies [25] and different properties of polymers [48-49] and molecules [50-52] using the essence of ML models. Applying ML methods in materials science accelerates the prediction of new novel materials [31, 53-54] which not only can bypass the criticality of the first principle calculations [55] having limitation to the requirement of advanced technical infrastructures but also the involved experimental synthesis and characterization of new materials.

Heusler alloys (HAs) are one of the most important classes of materials due to their versatile properties, which have major technical importance in spintronics research. They show half-metallic ferromagnetism [56], antiferromagnetism and compensated ferrimagnetism [57-58], superconductivity [59], shape memory effect [60], giant magnetoresistance [61], tunneling magnetoresistance [62], anomalous that helps to automate analytical model building and thus can take decision by finding the pattern in the data, without the intervention of human brains [73, 74], i.e. self-learning from data and experience. It is having the intersection of mathematics, statistics and computer science. ML methods are broadly classified into three types: a) supervised learning, having the labelled training data, b) unsupervised learning, with the unlabelled training data and c) reinforcement learning, with the intermix of training and testing phases [75]. In MI, majorly supervised learning methods are used. Previously, few studies [76-80] report about the use of ML algorithms to predict few HA properties. In this work, we have used supervised learning algorithm to predict structural and magnetic properties of all the classes of HAs, such that it is more detailed and general as compared to the previous reports. In section 2, methodology of general ML technique is described in step-by-by process; section 3 covers the results and discussion and section 4 ends with the conclusion.Hall effect [63], magneto-optic phenomena [64], thermoelectric effect and topological insulating behavior [65], magnetic semiconductivity [66], spin-gapless semiconductivity [67] etc. From application perspectives, they are suitable for spin-injection devices [68], magnetic tunnel junctions [62], spin-voltage generators [69], spin-valves [70], magnetic refrigerators [71], thermoelectric generators [65, 72] and many more. With their profound scientific and technological importance, search for new HAs is still going on. Thus accelerating the discovery of new HAs may help the materials research community to overcome several technological challenges. This is where ML can be a tool to predict new HAs based on the databases of existing current research on its domain. ML is a mathematical data analysis technique

## 2. Methodology

Prediction of the dependent or response variables with ML techniques generally follow the steps, shown in Fig. 1.

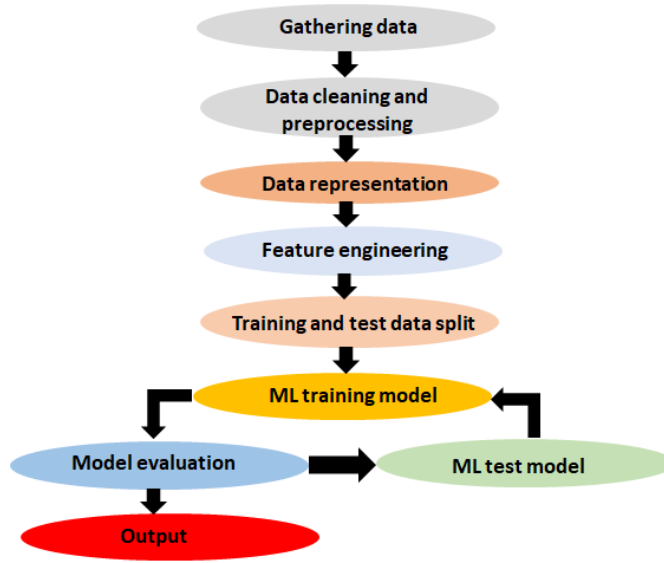

**Fig. 1.** Schematic diagram of general machine learning method.

2.1. Collection of data

The first step in any ML based approach is gathering data either from proprietary sources or from existing databases. The role of volume and quality of data is very crucial in any ML model. In materials sciences, popular sources of data can be found at refs. [26-30, 31], where information about materials' crystal structure and other properties is available. Specific to HAs, the database can be found at ref [30]. In this work, this database [30] is mainly used to predict the lattice constant, saturation magnetic moment and formation energy of full HAs, half HAs and inverse HAs. For quaternary HAs, a modular database has been prepared by using existing literatures [81-135].

2.2. Data cleaning and preprocessing

Data cleaning and preprocessing is the most crucial and time-consuming step in ML models. It is required to make the collected raw data into understandable format by removing incorrect, irrelevant and duplicate data and thus improving the productivity and efficiency of the model. This step also handles missing and noisy data. Outliers in a dataset degrades a model's performance metrics and requires special attention for its removal. There are few graphical (e.g. box plots) and numerical (e.g. Z-score) techniques which can give information about the outliers. Fig. 2 shows the box plots of the datasets for different types of HAs. Data which are beyond the minimum values of the boxes, are the potential outliers, which are removed here in prior before applying the ML model to the final dataset.

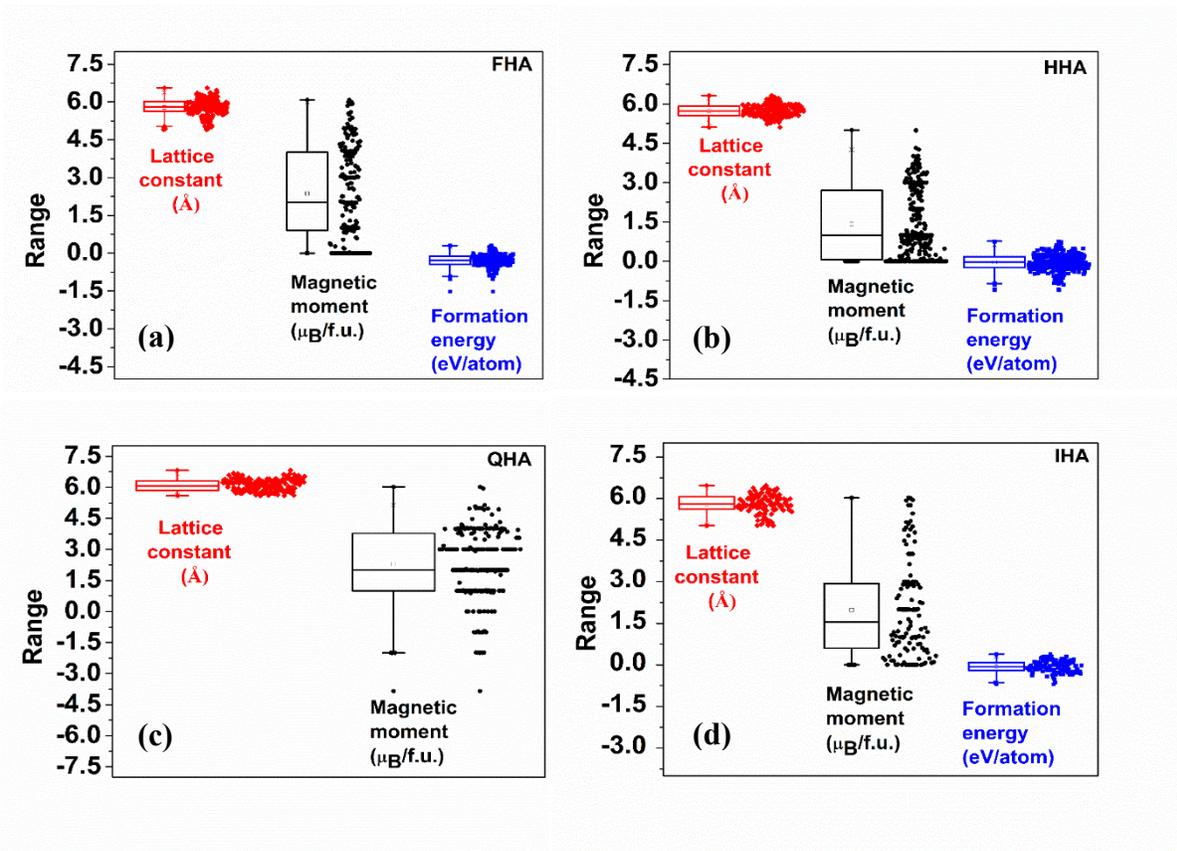

**Fig. 2.** Box plots for different types of HAs dataset **(a)** full HA, **(b)** half HA, **(c)** quaternary HA, **(d)** inverse HA.

2.3. Data visualization and feature engineering

Data visualization makes the data more understandable and interpretable by showing the trends and patterns, thus making the decision faster. There are different types of graphical data visualization tool e.g. histogram plots, scatter plots, bar charts, heat maps etc. Fig. 3 to Fig. 6 shows the scatter plots of the response variables: lattice constants, magnetic moment and formation energies for full HAs, half HAs, inverse HAs and quaternary HAs respectively. In ML, generally data is assumed to be distributed normally, to ease the model building, though it is not mandatory. Thus it is useful to check whether the data is normally distributed or not. Here, in this work,

Shapiro-Wilk (S-W) test [136] is used to confirm a nomally distributed data. In S-W test, if the p-value is greater than a significance level ($\alpha = 0.05$), then it is considered to be normally distributed, otherwise not.

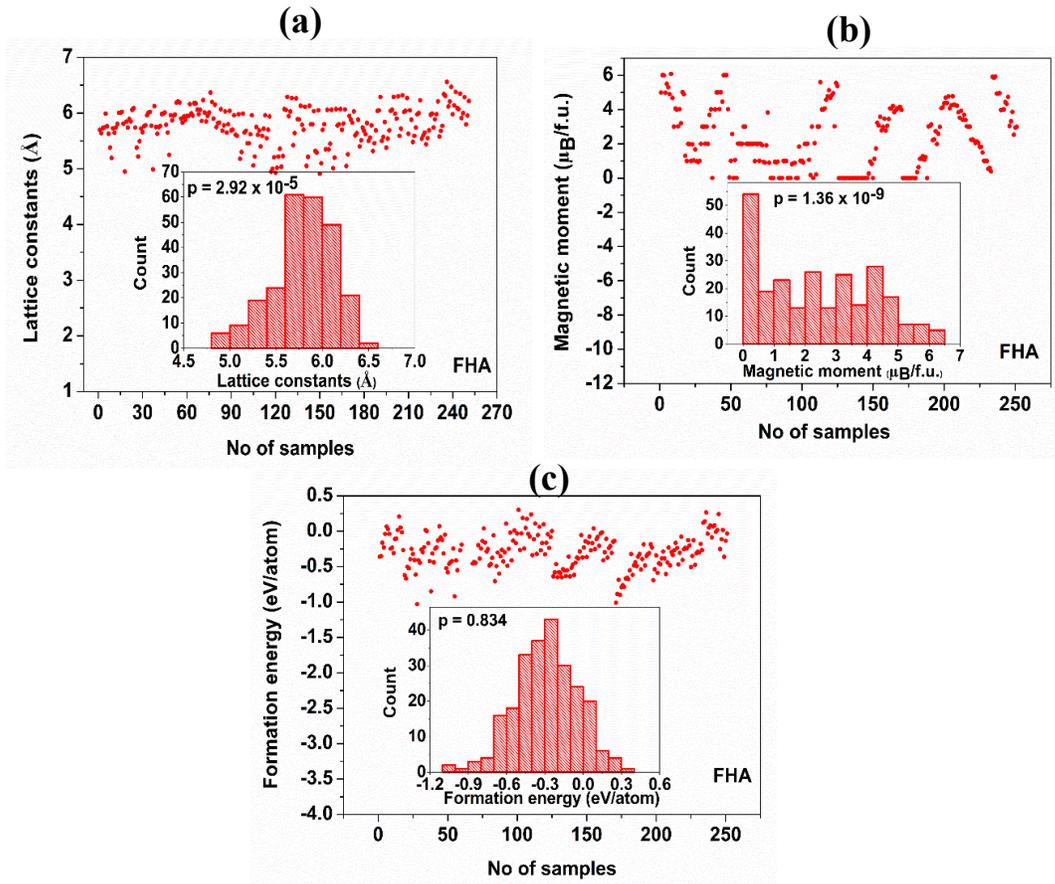

**Fig. 3.** Scatter plots of (a) lattice constant, (b) magnetic moment and (c) formation energy for full Heusler alloys. Inset shows the histogram plots of the same.

Inset curves in Figs. 3 (a) to 3(c) shows the histogram plots of each of the response variables in case of full HAs. From the curves it is evident that except formation energy (p =0.834), lattice constant (p = 2.92 x $10^{-5}$) and magnetic moment (p = 1.36 x $10^{-9}$) values are not normally distributed.

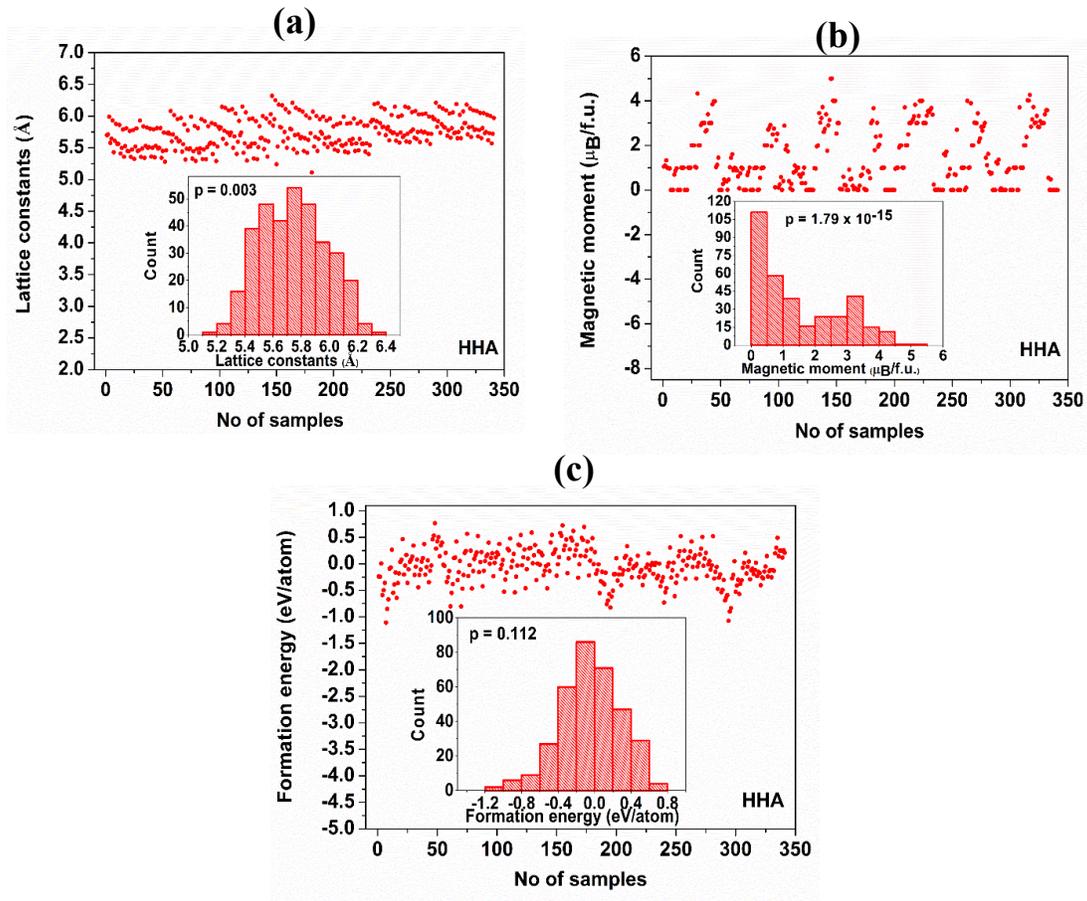

**Fig. 4.** Scatter plots of (a) lattice constant, (b) magnetic moment and (c) formation energy for half Heusler alloys. Inset shows the histogram plots of the same.

Figs. 4 (a) to (c) shows the scatter plots of the response variables for half HAs, along with their histograms. Lattice constant and magnetic moment with their p-values of 0.003 and 1.79 x $10^{-15}$ respectively shows that they are not normally distributed; however the formation energy values are normally distributed (p = 0.112).

Figs. 5 (a) to (c) shows that only formation energy values of inverse HAs are normally distributed with p = 0.27. Similarly, fig. 6 (a) and (b) shows that none of the lattice constant (p = 3.67 x $10^{-5}$) and the magnetic moment (p = 9.32 x $10^{-7}$) values of quaternary HAs are normally distributed.

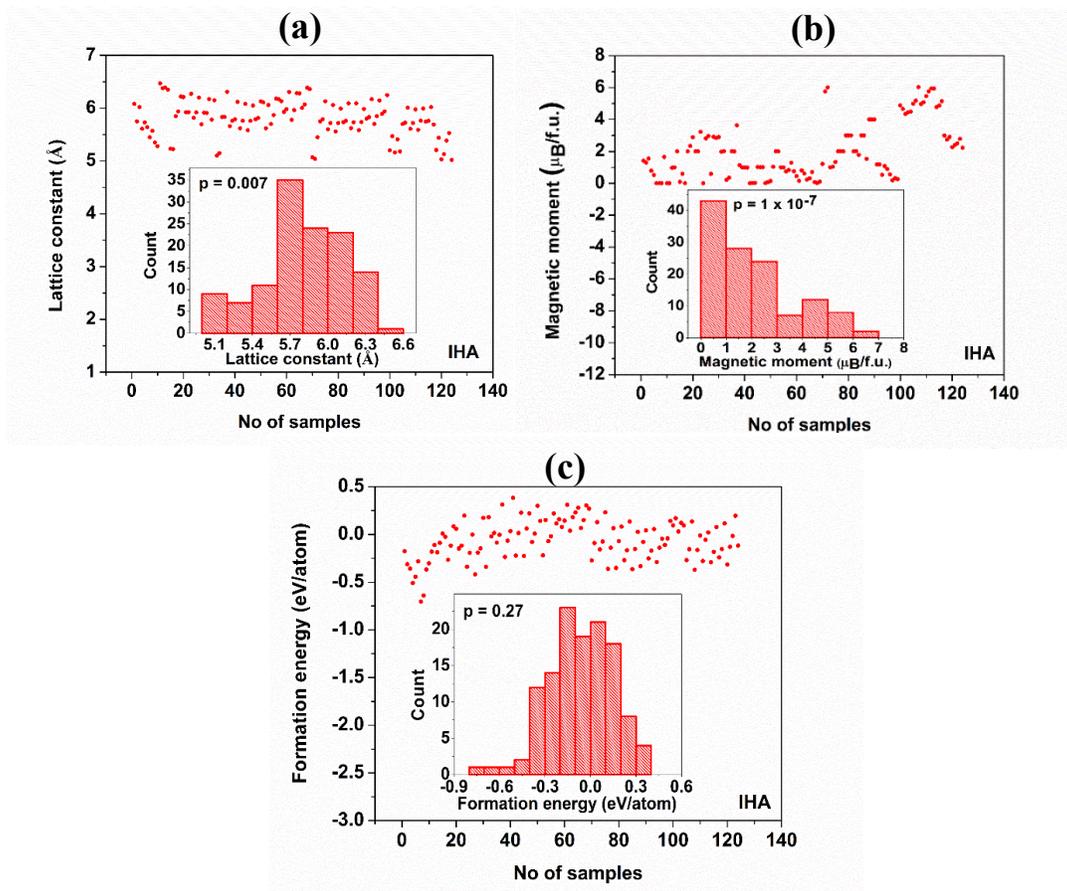

**Fig.5.** Scatter plots of (a) lattice constant, (b) magnetic moment and (c) formation energy for inverse Heusler alloys. Inset shows the histogram plots of the same.

Thus, from the scatter plots and the histogram plot it is evident that the data points of the response variables, except formation energy, are not normally distributed, and care is taken to counteract this observations while selecting the ML model to achieve best prediction results. To predict the response variables selecting the best feature variables is crucial and this is done by proper feature engineering techniques. In this work, feature variables are selected as: total number of valence electrons, ionic radii and electronegativities of the atomic elements of the HAs. The integrity of the features are filtered by prior variance thresholding techniques. Correlations among the input features can be found from correlation matrices shown in Figs. 8(d), 9 (d), 10 (d) and 11(d).

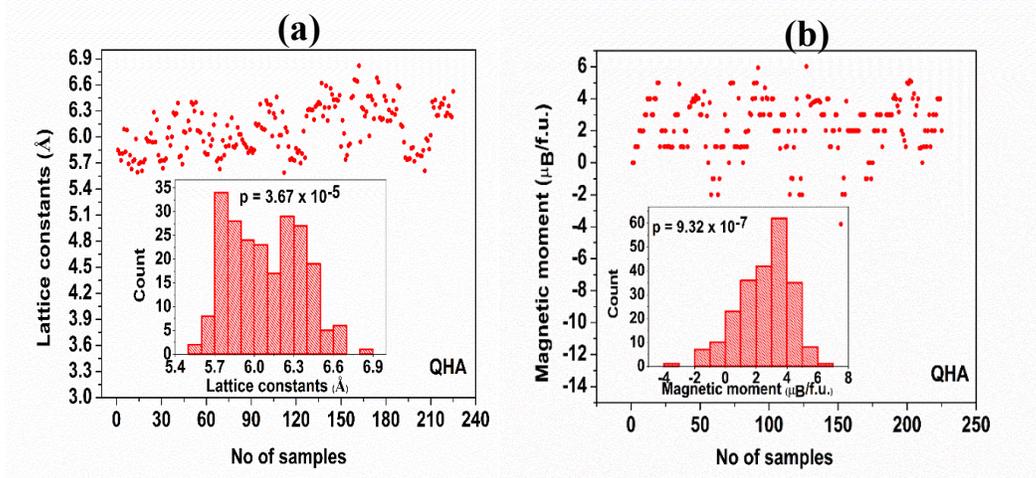

**Fig.6.** Scatter plots of (a) lattice constant, (b) magnetic moment for quaternary Heusler alloys. Inset shows the histogram plots of the same.

2.4. Machine learning model: Random forest

Selecting the best model is one of the most vital step in ML applications. Though it is not possible to select a single model for all types of uses, it saves both time and cost by choosing the right model for specific applications. The selection criteria may be depending upon the size of the training data and training time, inference time, dimensionality of the data and interpretability of the model. Also, it becomes more helpful when the model is sufficiently robust against multicollinearity, overfitting and outliers.

Random forest (RF) regression algorithm [137-139] with Python scikit-learn package [140] is used in this work to predict the structural and magnetic properties of the HA families. RF is one of the decision tree based ensemble approach which is famous for its robustness to overfitting and and outliers [141]. It works well for high dimensional, imbalanced data and moderately robust against the multicollinearity [142] between the predictor variables in the dataset, as it uses bootstrap sampling and feature sampling by randomly picking up different groups of features for different decision models, where each model selects a different sub-set of data points. RF can be considered as a variance reduction method, which can provide substantial improved results over a single decision tree based approach and may be used where the dataset has error rate considerably depending upon the variance [138]. Relative feature importance can also be predicted while using RF [141, 143-144] algorithms. RF does not require the data points to be normally distributed, which is generally assumed in linear regression based models.

2.5. Performance metrics

The ML model is trained with 70% of the dataset and the performance of the model is evaluated on rest of the 30% of the data, using cross-validation technique. Measuring the performance of the RF regression model is usually achieved using the coefficient of determination ($R^2$) (equation 1).

$$R^2 = 1 - \frac{RSS}{TSS} = 1 - \frac{\sum(x_i - \hat{x})^2}{\sum(x_i - \bar{x})^2} \qquad (1)$$

where, RSS = sum of square of residuals and TSS = total sum of squares, $x_i$ = ith observation for the feature x, and $\bar{x}$ is the mean value of x in the dataset.

$R^2$ can be taken as a statistical measure of the goodness of fit between the ML-predicted and DFT-computed results and gives information about the extent of correlation between them. However, adjusted $R^2$ (equation 2) gives more accurate measure of the correlation by considering all other independent variables, thus making the model more robust against adding new feature variables.

$$\text{Adj. } R^2 = 1 - \frac{(1-R^2)(N-1)}{N-p-1} \qquad (2)$$

where, N = total # of observations and p = # of independent variables.

We have used Adj. $R^2$ as the performance metric in this work.

### 3. Results and discussion

The crystal structure of different types of HAs is shown in Fig. 7, in which full HAs (FHA), half HAs (HHA), inverse HAs (IHA) and quaternary HAs (QHA) are symbolized as in Table 1, where X, X´, Y atoms are transition metal atoms and Z is the main group element. Table .1 shows the crystal structure details of different types of HAs with their symbols, space groups, Wyckoff positions and site occupancies with reference to Fig. 7.

**Table 1** Crystal structure details of different types of HAs

| Type | Symbol | Space group | Wyckoff positions | Sites A | B | C | D |
|------|--------|-------------|-------------------|---|---|---|---|
| FHA | $X_2YZ$ | Fm-3m (225) | X: 8c (1/4,1/4,1/4) and (3/4/,3/4,3/4) | X | Y | X | Z |
|  |  |  | Y: 4b (1/2,1/2,1/2) |  |  |  |  |
|  |  |  | Z: 4a (0,0,0) |  |  |  |  |
| HHA | XYZ | F-43m (216) | X: 4a (0,0,0) | X | Y | Void | Z |
|  |  |  | Y: 4b (1/2,1/2,1/2) |  |  |  |  |
|  |  |  | Z: 4c (1/4,1/4,1/4) |  |  |  |  |
| QHA | XX'YZ | F-43m (216) | X: (3/4,3/4,3/4) | X | Y | X' | Z |
|  |  |  | X': (1/4,1/4,1/4) |  |  |  |  |
|  |  |  | Y: (1/2,1/2,1/2) |  |  |  |  |
|  |  |  | Z: (0,0,0) |  |  |  |  |
| IHA | $X_2YZ$ | F-43m (216) | X: 4a (0,0,0) and 4d (3/4/,3/4,3/4) | X | X | Y | Z |
|  |  |  | Y: 4b (1/2,1/2,1/2) |  |  |  |  |
|  |  |  | Z: 4c (1/4,1/4,1/4) |  |  |  |  |

In this work, prediction of the structural and magnetic properties of all types of HAs is carried out by taking the seven input features as: $N_v$ = Total # of valence electrons; r (i) = Ionic radius and E (i) = Electronegativity of i atom, where i = X, X´, Y and Z;

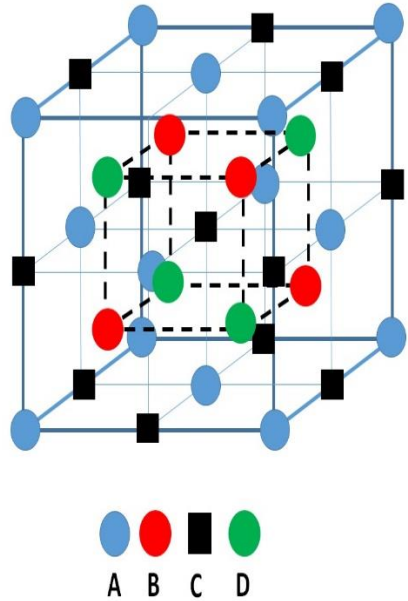

**Fig.7.** Schematic diagram of crystal structure with different sites of occupancy for different types of Heusler alloys [145].

Fig. 8 (a) to (c) show the parity plots of the lattice constant, magnetic moment and formation energies of full HAs indicating the variation of the machine learning predicted properties with the DFT- computed values. The inset shows the importance of each input feature variable in predicting the results. Close agreement of the model prediction performance can be concluded from the adjusted $R^2$ values, which is greater than 0.8 in each case. Fig. 8 (d) shows the correlation matrix explaining the mutual correlations among the feature variables for FHAs. It can be seen that even though there is moderate correlations between the input feature variables, RF model can predict the response variables with sufficient accuracy.

Similar parity plots of the lattice constant, magnetic moment, formation energies of half HAs, and inverse HAs are shown in Fig. 9 and Fig. 10 respectively, where the prediction is reasonably well. Fig. 11 shows the parity plots of (a) lattice constant and (b) magnetic moment for quaternary Heusler alloys. In each plots, the inset shows the importances of the input features and it is observed that for almost every cases, the role of the valence electrons number is important. It can also be noticed that the performance of the RF regression model is not hampered by the moderate correlations present between some of the prediction variables, as observed from the correlation matrices Figs.8 (d), 9 (d), 10 (d) and 11 (d).

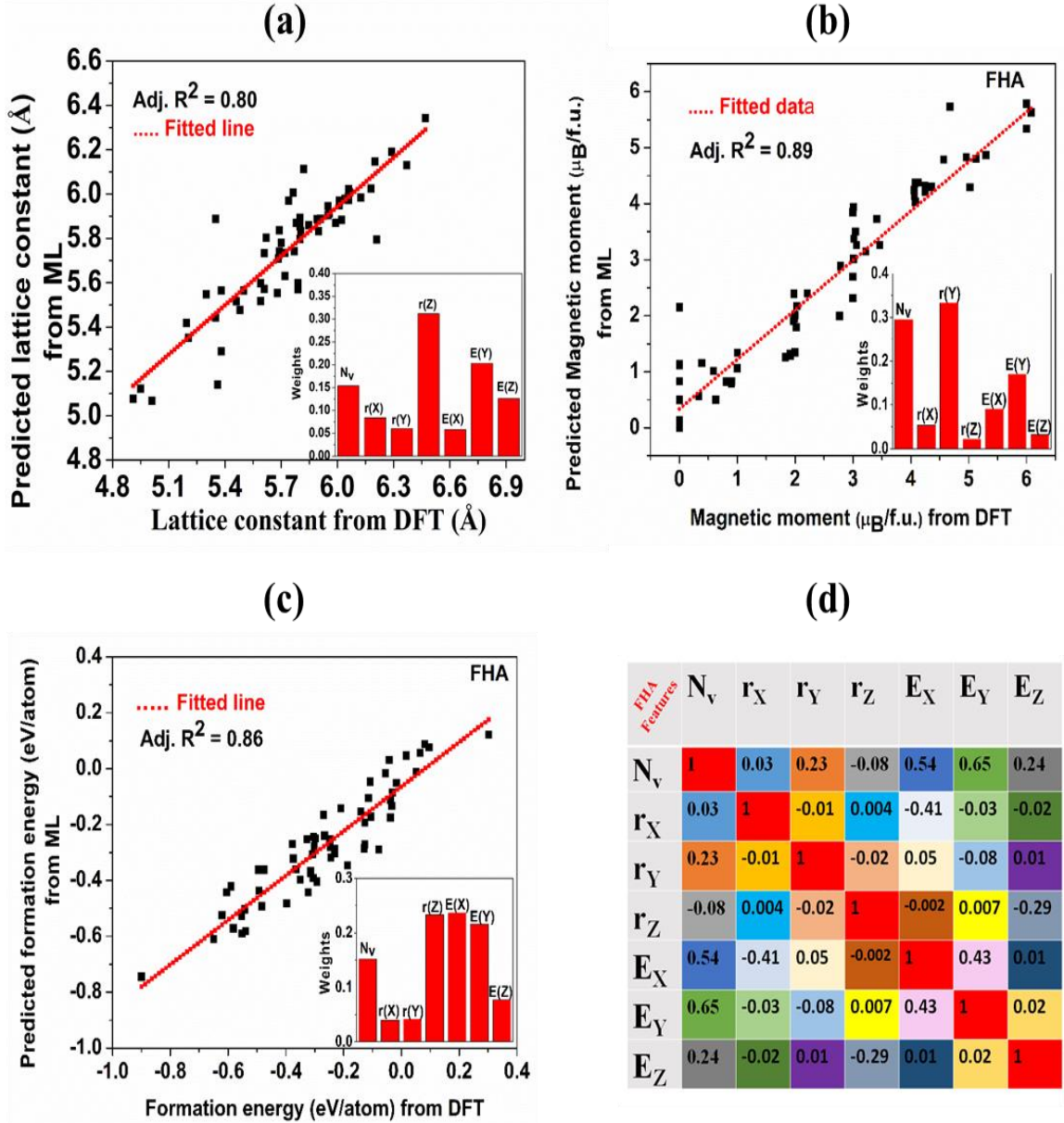

**Fig.8.** Parity plots of the response variables **(a)** lattice constant, **(b)** magnetic moment, **(c)** formation energy for full Heusler alloys showing the variation of machine learning predicted values against the values computed from density functional theory. **(d)** Correlation matrix showing the correlation between the predictor feature variables used to predict the above response variables.

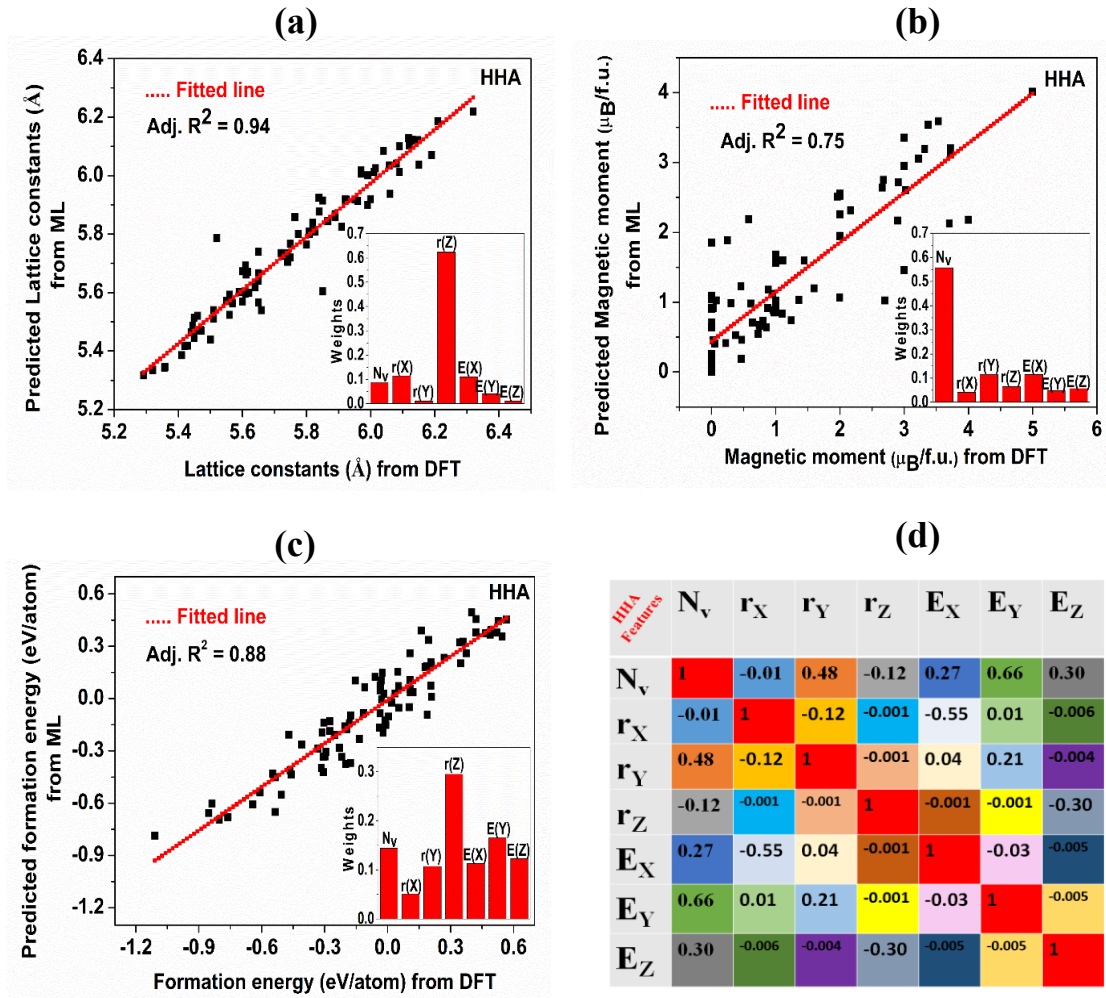

**Fig. 9.** Parity plots of the response variables **(a)** lattice constant, **(b)** magnetic moment, **(c)** formation energy for half Heusler alloys showing the variation of machine learning predicted values against the values computed from density functional theory. **(d)** Correlation matrix showing the correlation between the predictor feature variables used to predict the above response variables.

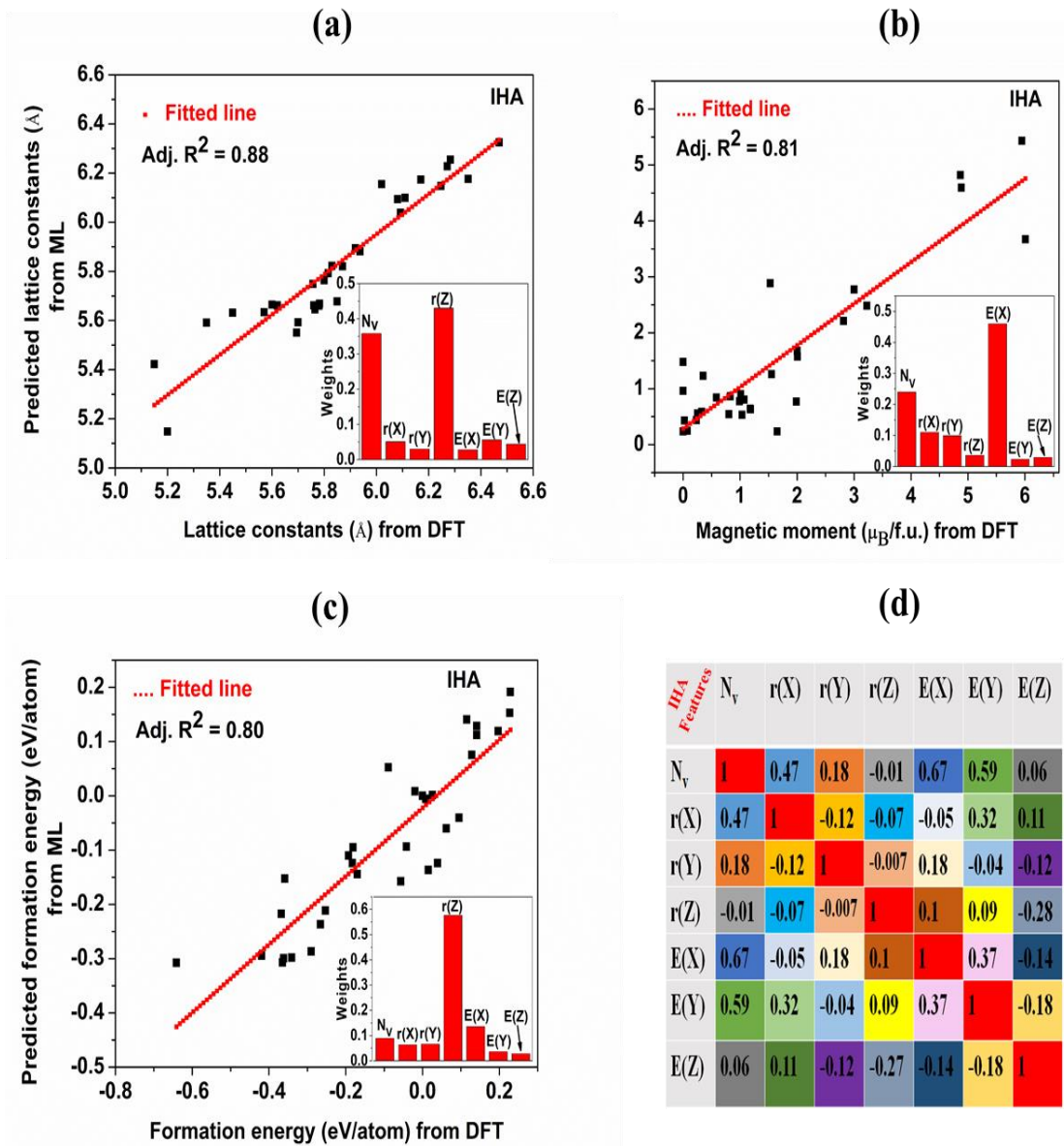

**Fig. 10.** Parity plots of the response variables **(a)** lattice constant, **(b)** magnetic moment, **(c)** formation energy for inverse Heusler alloys showing the variation of machine learning predicted values against the values computed from density functional theory. **(d)** Correlation matrix showing the correlation between the predictor feature variables used to predict the above response variables.

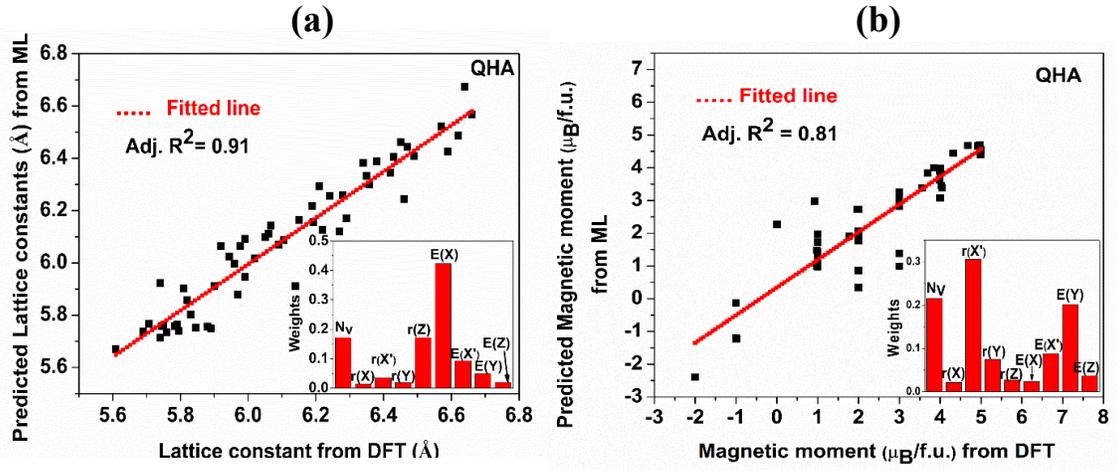

**Fig. 11.** Parity plots of the response variables **(a)** lattice constant, **(b)** magnetic moment, for quaternary Heusler alloys showing the variation of machine learning predicted values against the values computed from density functional theory. **(c)** Correlation matrix showing the correlation between the predictor feature variables used to predict the above response variables.

**Case study:**

Based on the model analysis and prediction results above, a case study is performed by selecting two samples: $Co_2FeGe$ full HA and CoFeTiSi quaternary HA. Both these alloys were studied using density functional theory (DFT) based first principle calculations and were also synthesized experimentally in our previous works [146-147]. The comparison table of the ML predicted results and the results obtained from theoretical calculations and experimental characterizations of the prepared samples is described in Table 2.

**Table 2** Comparison table showing the values of lattice constant and saturation magnetic moment of $Co_2FeGe$ full HA and CoFeTiSi quaternary HA, obtained from theory, experiments and machine learning based predictions.

| Heusler Alloys | Theory (DFT calculations) | | Experiments | | Machine learning predictions (This work) | |
|---|---|---|---|---|---|---|
| | Lattice constant (Å) | Magnetic Moment ($\mu_B$/f.u.) | Lattice constant (Å) | Magnetic Moment ($\mu_B$/f.u.) | Lattice constant (Å) | Magnetic Moment ($\mu_B$/f.u.) |
| $Co_2FeGe$ (FHA) [146] | 5.74 | 5.69 (GGA) 5.99 (GGA+U) | 5.74 | 6.1 | 5.78 | 5.64 |
| CoFeTiSi (QHA) [147] | 5.74 | 1 | 5.74 | 1.24 | 5.78 | 1.03 |

Table 2 shows that the performance of Random Forest model used in this work in predicting the values of lattice constant and saturation magnetic moment is at par with the results obtained from both theory and experiments and thus can be treated as an reliable alternative way to find the HA properties.

### 4. Conclusion

Systematic process is followed to predict structural and magnetic properties of different types of HAs, using machine learning technique. Existing databases are used to predict the properties of full HAs, half HAs and inverse HAs; however, an indigenous database is created with the lattice constant, magnetic moment of quaternary HAs from the existing literatures. Decision tree based random forest regression algorithm is used to predict the response variables: lattice constant, magnetic moment and formation energies, after closely visualizing the input datasets, which is majorly not normally distributed as per Shapiro-Wilk test. This non-parametric approach is robust against overfitting and outliers and performs well in predicting the response variables even though some of the input features have moderate correlations between them. The outcome of the Random Forest model is reasonably accurate as observed from the adjusted $R^2$ values, which remains in the range of 0.8-0.94 for most of the predictions. Feature importance computations show that number of valence electrons remains one of the most important input features for the predictions. Use of Random Forest model in predicting the mentioned response variables for all types of HAs is carried out for the first time to the best of our knowledge. Case studies were mentioned comparing the machine learning predicted results with the earlier theoretical calculations and experimental observations, suggesting the high accuracy of the Random Forest model. This work will be very helpful to the materials researchers such that a preliminary knowledge of the most important structural and magnetic properties of any types of unknown HAs can also be accurately predicted even before any first principle based calculations and experimental synthesis of the alloys.

## Acknowledgements

Mr. Srimanta Mitra wish to thank Dr. B. Narasiha Sharma (Head, SSD/EOSDIG/SEDA), Mrs. Arti Sarkar (Group Director, EOSDIG/SEDA), Mr. Soumya S. Sarkar (Deputy Director, SEDA), Mr. N. M. Desai (Director), Space Applications Centre, ISRO, Ahmedabad for providing their support and motivations to complete this work.